\documentclass{eptcs}
\pdfoutput =1
\providecommand{\event}{GandALF 2021}
\usepackage{amsfonts}
\usepackage{amsmath}
\usepackage{amsthm}
\usepackage{amssymb}
\usepackage[utf8]{inputenc}
\usepackage[english]{babel}

\usepackage{scalerel,stackengine}
\stackMath
\newcommand\reallywidehat[1]{%
\savestack{\tmpbox}{\stretchto{%
  \scaleto{%
    \scalerel*[\widthof{\ensuremath{#1}}]{\kern-.6pt\bigwedge\kern-.6pt}%
    {\rule[-\textheight/2]{1ex}{\textheight}}
  }{\textheight}%
}{0.5ex}}%
\stackon[1pt]{#1}{\tmpbox}%
}
\theoremstyle{definition}
\newtheorem{definition}{Definition}[section]

\theoremstyle{plain}
\newtheorem{theorem}{Theorem}[section]
\newtheorem{lemma}{Lemma}[section]
\newtheorem{proposition}{Proposition}[section]
\newtheorem{corollary}{Corollary}[section]

\title{Finite Model Property and Bisimulation for LFD}
\author{Raoul Koudijs,\\ ILLC Amsterdam}
\def\titlerunning{Finite Model Property and Bisimulation for Local Logics of Dependence}
\def\authorrunning{R.S. Koudijs}
\begin{document}
\maketitle

\begin{abstract}
Recently, Baltag \& van Benthem \cite{BaltagvBenthemLFD} introduced a new decidable logic of functional dependence (LFD) with local dependence formulas and dependence quantifiers. The language is interpreted over dependence models, which are pairs of first-order structures with a set of available variable  assignments, also called a team.  The team associated with a dependence model can be seen as a labelled transition system over which LFD becomes a modal logic, where the dependence quantifiers become modalities and local dependence formulas are treated as special atoms. In this paper, we introduce appropriate notions of bisimulation characterizing LFD (and some related logics) as a fragment of first order logic (FOL), and show it is equivalent to a notion of bisimulation along more standard lines proposed in \cite{local_deps}, yet more efficient for bisimilarity-checking. Our main result is that LFD has the finite model property (FMP), by a new application of Herwig's theorem on extending partial isomorphisms \cite{Herwig1998ExtendingPI}.
\end{abstract}

\section{\textbf{Introduction}}
Recently, Baltag \& van Benthem introduced a logic of functional dependence LFD, which can be seen both as a first-order and a modal logic. As a first-order language, it is a fragment of relational FOL with dependence quantifiers and complex local dependence formulas. As a modal language, it extends classical propositional calculus with dependence modalities and atomic local dependence formulas. We are interested in the concept of dependence between \textit{variables}, so statements of the form "the variable $y$ depends on the variables $X$". In many logics of (in)dependence, formulas are evaluated at a team (i.e. a set
of variable assignments). By contrast, LFD is interpreted over \textit{dependence models}, which are pairs $(M,A)$ of a first-order structure $M$ with a fixed team $A\subseteq M^V$ ($V$ some set of variables), and formulas are evaluated at individual assignment $s\in A$ from this fixed team.  Consequently, LFD models the more basic concept of local dependence: a variable $y$ depends on the set $X$ \textit{locally at} $s$ if for all assignments $t\in A$; if $s(x)=t(x)$ then $s(y)=t(y)$. Then global dependence on the whole team $A$ becomes definable as local dependence at all assignments in $A$. Underlying this local perspective is the equivalent \textit{modal} semantics for LFD, where the assignments become abstract states in a transition system and the dependence quantifiers become modalities.\newline\newline
LFD was shown to be \textit{\textit{decidable}} \cite{BaltagvBenthemLFD} by proving completeness and the finite model property w.r.t syntactic type-models that resemble the 'quasi-models' studied in connection with the Guarded Fragment \cite{Exploring_Lo_Dyn}\cite{ModLangBoundFragmPredLog}. The question whether LFD has the FMP w.r.t. dependence models was left as an open problem \cite{BaltagvBenthemLFD}. In this paper, we propose suitable notions of bisimulation characterizing LFD and some related local dependence logics. Independently, other notions of bisimulation for these logics, along more standard lines have been proposed in \cite{local_deps}. We show that our notions are equivalent to these, but more efficient for bisimilarity-checking. The main result of this paper is that LFD has the FMP (w.r.t. dependence models) by a new application of Herwig's theorem on extending partial isomorphisms. Our proof strategy is similar to the proof of the FMP for the Guarded fragment (GF) in \cite{grAdel_1999}, but we exploit more conditions guaranteed by Herwig's theorem. Moreover, we present a modal proof of the FMP in the appendix, using a generalised version of Herwig's theorem which additionally guarantees the omittance of packed homomorphic images.\newline\newline
We first introduce the language LFD, the dependence model-semantics and the non-standard type model-semantics. A pair $(V,\tau)$, where $V$ is set of variables and $\tau$ is a relational language is called a \textit{vocabulary}. When both $V$ and $\tau$ are finite, we say that $(V,\tau)$ is a finite vocabulary. We write $FOL[V,\tau]$ for the set of first-order formulas with variables in $V$ (both free and bound) and predicates in $\tau$, and similarly for $LFD[V,\tau]$. We assume that each vocabulary becomes equipped with an arity map $\mathrm{ar}:\tau\to\mathbb{N}$.

\begin{definition}{(\textbf{Syntax})} The language $LFD[V,\tau]$ is recursively defined by:
\[\varphi::= P\mathbf{x}\;|\;\neg\varphi\;|\;\varphi\wedge\varphi\;|\;\mathbb{D}_X\varphi\;|\;D_Xy\]
where $X\subseteq V$ is a \textit{finite} set of variables, $y\in V$ an individual variable, $P\in\tau$ a predicate symbol and $\mathbf{x}=(x_1,...,x_n)\in V^{ar(P)}$ a finite string of variables.\footnote{LFD as a modal language is generated by the same definition, but where $D_Xy(\;),P\mathbf{x}(\;)$ become unary predicates in $\tau$. Fixing notation, for any $Y\subseteq V$, we write $s\models D_XY$ if $s\models D_Xy$ holds for all $y\in Y$. We also skip the set brackets for singletons, writing $D_xY$ for $D_{\{x\}}Y$ , and $D_xy$ for $D_{\{x\}}\{y\}$.} For every $\varphi\in LFD$, we define its free variables by:
\begin{itemize}
    \item $free(Px_1...x_n)=\{x_1,...,x_n\}$
    \item $free(D_Xy)=free(\mathbb{D}_X\varphi)=X$
    \item $free(\neg\varphi)=free(\varphi)$,\; $free(\varphi\wedge\psi)=free(\varphi)\cup free(\psi)$
\end{itemize}
\end{definition}

\vspace{0.015in}

\begin{definition}{(\textbf{Dependence Models})} A \textit{dependence model} (for the vocabulary $(V,\tau)$) $\mathbb{M}$ is a pair $\mathbb{M}=(M,A)$ of a relational first-order $\tau$-structure $M=(O, I)$ (with domain $O$ and interpretation map $I$), together with a team $A\subseteq O^V$.\footnote{We will use capital letters $M=(O,I)$ for first-order structures and blackboard bold capital letters $\mathbb{M}=(M,A)-((O,I),A)$ for dependence models.} For each $X\subseteq V$, we have an agreement relation $=_X$ on assignments from the team $A$: $s=_Xt$ iff $s\restriction X=t\restriction X$.\footnote{In fact, only the relations $=_X$ for \textit{finite} $X\subseteq V$ are relevant to the semantics} A dependence model $(M,A)$ is called \textit{distinguished} if distinct variables can only take distinct values\footnote{It is helpful to think of the objects in distinguished dependence models as being typed (pr labelled) by a unique variable.} and \textit{full} if $A=O^V$.
\end{definition}

\begin{definition}{(\textbf{Semantics})} Truth of a formula $\varphi$ in a dependence model $\mathbb{M}=(M,A)$ at an assignment $s\in A$ is defined by the following clauses  (the Boolean cases are defined as usual):\footnote{We drop the model-index $\mathbb{M}$ if it is understood by context.}
\begin{align*}
& s\models P\mathbf{x}\;\textrm{iff}\;s(\mathbf{x})\in I^{\mathbb{M}}(P)\\
& s\models\mathbb{D}_X\varphi\;\textrm{iff}\;t\models\varphi\;\textrm{holds for all}\;t\in A \;\textrm{with}\; s =_X t\\
& s\models D_Xy\;\textrm{iff}\;s=_X t\;\textrm{implies}\; s=_yt\;\textrm{for all}\;t\in A.
\end{align*}
\end{definition}
Where $s(\mathbf{x})$ denotes the tuple $(s(x_1),...,s(x_m))$ if $\mathbf{x}=(x_1,...,x_m)$. LFD (over finite vocabularies) also has a non-standard type model-semantics for which it is weakly complete (LFD is strongly complete w.r.t dependence models \cite{BaltagvBenthemLFD}). For $\psi\in LFD$, let $V_{\psi}$ be the set of 'relevant variables' \textit{occurring} in $\psi$. Add to $\{\psi\}$ all formulas $D_Xy$ for $X\cup\{y\}\subseteq V_{\psi}$. Close the resulting set under subformulas, as well as one round of negations, where explicit negations are left as they are. The resulting finite set $\Psi:=Cl(\{\psi\})$ is called the \textit{closure} of $\psi$.\footnote{Note that $\Psi$ belongs to LFD over the \textit{finite vocabulary} $(V_{\psi},\{P\;|\;P\;\textrm{occurs in}\;\psi\})$, so type models are only defined for LFD over finite vocabularies.}

\begin{definition}{($\Psi$-\textbf{Types})} Let $\Psi=Cl(\{\psi\})$ be a closure with relevant variables $V_{\psi}$ (i.e. the variables occurring in $\Psi$). A subset $\Sigma\subseteq\Psi$ is a $\Psi$-\textit{type} if it satisfies the following conditions (where all formulas mentioned run over $\Psi$ only):
\begin{itemize}
    \item [(a)] $\neg\psi\in\Sigma$ iff $\psi\not\in\Sigma$\qquad\qquad\qquad\qquad\qquad \;\;(d) $D_Xx\in\Sigma$ for all $x\in X$
    \item [(b)] $(\psi\wedge\chi)\in\Sigma$ iff $\psi\in\Sigma$ and $\chi\in\Sigma$\qquad\qquad \;\;(e) if $D_XY, D_YZ\in\Sigma$, then
    \item [(c)] if $\mathbb{D}_X\psi\in\Sigma$, then $\psi\in\Sigma$
\end{itemize}
\end{definition}
For each $X\subseteq V_{\psi}$, we define a relation $\sim_X$ on $\Psi$-types as follows:
\begin{align*}
\Sigma\sim_X\Delta\qquad\textrm{iff}\qquad&\{\phi\in\Sigma\;|\;\phi\in\Psi\;\&\;free(\phi)\subseteq D^{\Sigma}_X\}=\{\phi\in\Delta\;|\;\phi\in\Psi\;\&\;free(\phi)\subseteq D^{\Sigma}_X\}
\end{align*}
where $D^{\Sigma}_X=\{y\in V_{\varphi}\;|\;D_Xy\in\Sigma\}$ is the dependence-closure of $X$ w.r.t and $\Sigma$. Observe that $\Sigma\sim_X\Delta$ implies $D^{\Sigma}_X=D^{\Delta}_X$ since $free(D_Xy)=X$. 

\begin{definition}{(\textbf{Type Models})} A \textit{type model} (for $\Psi$) is a family of $\Psi$-types satisfying:
\begin{itemize}
    \item if $\mathbb{E}_X\psi\in\Sigma\in\mathfrak{M}$, then there exists a $\Delta\in\mathfrak{M}$, such that $\psi\in\Delta$ and $\Sigma\sim_X\Delta$.
    \item $\Sigma\sim_{\emptyset}\Delta$ holds for all $\Sigma,\Delta\in\mathfrak{M}$.
\end{itemize}
Type models are always finite because closure sets $\Psi$ are. This proves the decidability of LFD as the logic is (weakly) complete w.r.t. type models.\footnote{In \cite{BaltagvBenthemLFD} it is shown that a generated submodel of the canonical model is a type model, establishing weak completeness of LFD w.r.t type models.} The semantics is given by "truth = membership":
\[\Delta\models\psi\qquad\textrm{iff}\qquad\psi\in\Delta\]
\end{definition}

\subsection{\textbf{Tree Model Property}}
Clearly, for a given closure set $\Psi$, any dependence model $\mathbb{M}=(M,A)$ induces a type model by abstracting away from the concrete nature of the assignments. That is, whenever $type_{\Psi}(s):=\{\phi\in\Psi\;|\;s\models_{\mathbb{M}}\phi\}$ is well-defined (i.e. $M$ is a structure for the language of $\Psi$), the collection of all $\Psi$-types $\{type_{\Psi}(s)\;|\;s\in A\}$ is a type model. Conversely, Baltag \& van Benthem show that any type model for a given $\Psi$ can be represented as the set of $\Psi$-types of a dependence model, by a complicated unravelling construction \cite{BaltagvBenthemLFD}. To formulate this result, we first introduce the graph-theoretic notion of a $k$-tree (the definition is taken from \cite{grAdel_1999}). Say that an $r$-tuple of objects $\mathbf{a}$ from a $\tau$-structure $M$ is \textit{live} in $M$, if there is some $r$-ary $P\in\tau$ such that $M\models P\mathbf{a}$.

\begin{definition}{(\textbf{$k$-Tree})} A $\tau$-structure $M$ is a \textit{$k$-tree} if there exists a tree (i.e. an acyclic, connected graph) $T=(V,E)$ and a function $F:V\to\{X\subseteq M\;|\;|X|\leq k\}$, assigning to every node $v\in V$ of $T$ a set $F(v)$ of at most $k$ elements of $M$, such that the following two conditions hold.
\begin{itemize}
    \item [(i)] For every live tuple $\mathbf{a}=(a_1,...,a_r)$ from $M$, there is some node $v$ such that $\{a_1,...,a_r\}\subseteq F(v)$.
    \item [(ii)] For every element $a$ of $M$, the set of nodes $\{v\in V\;|\;a\in F(v)\}$ is connected (and hence induces a subtree of $T$).
\end{itemize}
$M$ is of finite branching degree if $T$ is, that is if the set of neighbours of every node in $T$ is finite.
\end{definition}

\begin{theorem}{(\textbf{Representation of Type Models})} \cite{BaltagvBenthemLFD} Let $\mathfrak{M}$ be a type model for $\Psi$ with relevant variables $V$ and $|V|=k$.\footnote{So if $\Psi=Cl(\{\psi\})$, then $V=V_{\psi}$.} Then there exists a dependence model $\mathbb{M}=(M,A)$ such that $\mathfrak{M}=\{type_{\Psi}(s)\;|\;s\in A\}$ and $M$ is a $k$-tree of finite branching degree.

\begin{proof}
Type models are always finite, so let $|\mathfrak{M}|=m$. The model is constructed as follows. Fix a $\Psi$-type $\Sigma_0\in\mathfrak{M}$, and define a \textit{good path} to be a finite sequence $\pi=\langle\Sigma_0,X_1,...,X_n,\Sigma_n\rangle$ of any positive length such that for each $i\leq n$ (a) $\Sigma_i\in\mathfrak{M}, X_i\subseteq V$ and (b) $\Sigma_{i-1}\sim_{X_i}\Sigma_i$. We write $last(\pi)=\Sigma_n$ for the last element of $\pi$, and $lh(\pi)=n+1$ for the length of this sequence (not counting the variable sets). For each good path $\pi$, we define the \textit{path assignment} $v_{\pi}$, assigning objects of the form $(\pi,x)$ to variables $x\in V$:
\begin{align}
    & v_{\pi}(x)=(\pi,x)\;\textrm{if}\;\pi\;\textrm{has length 1, i.e.}\;\pi=\langle\Sigma_0\rangle\;\textrm{is the root of our tree}.\\
    & v_{\pi}(x)=v_{\pi'}(x)\;\textrm{if}\;\pi=(\pi',X,\Sigma)\;\textrm{with}\;x\in D^{last(\pi')}_X\\
    & v_{\pi}(x)=(\pi,x)\;\textrm{if}\;\pi=(\pi',X,\Sigma)\;\textrm{with}\;x\not\in D^{last(\pi')}_X
\end{align}
where $D^{last(\pi')}_X=\{y\in V_{\varphi}\;|\;D_Xy\in last(\pi')\}$ is the \textit{dependence-closure} of $X$ w.r.t. the $\Psi$-type $last(\pi')$. So new objects are created whenever the value for a variable is not locally determined by the predecessor path. We obtain a team $A:=\{v_{\pi}\;|\;\pi\;\textrm{a good path}\}$, which is a team on the first-order structure $M=(O,I)$, where $O=\bigcup_{v_{\pi}\in A}v_{\pi}[V]$ is just the union of the ranges of all path assignments in $A$. Finally, the interpretation $I$ is given by the following ‘coherence condition’\cite{BaltagvBenthemLFD}:
\begin{quote}
    $I(P)$ holds for a finite sequence of objects $\overline{(\pi, x)}$ in $O$ if all paths $\pi$ occurring in the sequence are linearly ordered by the relation of initial segment, and the formula $P\mathbf{x}$ occurs in $last(\pi^*)$ on the longest path $\pi^*$ among these
\end{quote}
this yields a distinguished dependence model $\mathbb{M}=(M,A)$, as objects $(\pi,x)$ are \textit{typed} by a unique variable $x$. The set of all good paths ordered by initial segment forms a tree $T$ of finite branching degree (bounded by $|\mathcal{P}(V)|\times|\mathfrak{M}|=2^k\times m$), so that the map $F:\pi\mapsto v_{\pi}[V]$ witnesses the fact that $M$ is a $k$-tree of finite branching degree. Finally, the following \textit{truth lemma} \cite{BaltagvBenthemLFD} shows that $\mathfrak{M}=\{type_{\Psi}(s)\;|\;s\in A\}$:
\begin{lemma} For all formulas $\psi\in\Phi$ and good paths $\pi:\;\;$ $\mathbb{M},v_{\pi}\models\psi$ iff $\psi\in last(\pi)$
\end{lemma}
This is  because for every $\Delta\in\mathfrak{M}$, there is a good path $\pi_{\Delta}=\langle\Sigma_0,\emptyset,\Delta\rangle$ with $last(\pi_{\Delta})=\Delta$. Also, note that we are free to choose the initial fixed type $\Sigma_0$ from $\mathfrak{M}$ in the definition of good path.
\end{proof}
\end{theorem}

\begin{corollary}{\textbf{Tree Model Property}}\\
If $\psi\in LFD$ is satisfiable with $|V_{\psi}|=k$, there is a dependence model $\mathbb{M}=(M,A)$, where $M$ is $k$-tree of finite branching degree, satisfying $\varphi$ at the root assignment.
\end{corollary}

\begin{definition}{(\textbf{First-Order Translation})} Although interpreted over a generalised semantics, LFD over any vocabulary $(V,\tau)$ with $V$ finite can be encoded back into FOL over standard structures.\footnote{$CRS$-quantifiers recapture standard first-order quantification over full generalised assignment models, and for \textit{finite} $V$ the dependence quantifiers of LFD and $CRS$-quantifiers are inter-definable; for infinite $V$ the 2 notions seem to be independent \cite{BaltagvBenthemLFD}.} So let $V$ be finite with a fixed enumeration $\mathbf{v}=(v_1,...,v_n)$. We double the amount of variables, creating a set of copied variables $V'$ from the variables in $V$ and introduce a new $n$-ary predicate $A$ such that $A\mathbf{v}$ encodes the fact that the tuples of values assigned to $\mathbf{v}$ by the current assignment is the range of some admissible assignment from the team (this is a tuple because $V$ is finite). The first-order translation $tr:LFD[V,\tau]\to FOL[V\cup V',\tau]$ is defined by \cite{BaltagvBenthemLFD}:
\begin{itemize}
    \item $tr(P\mathbf{x}) = P\mathbf{x}$ and $tr$ commutes with Boolean connectives
    \item $tr(\mathbb{D}_X\psi) = \forall\mathbf{z}(A\mathbf{v}\to tr(\psi))$, where $\mathbf{v}$ is the enumeration of all the variables in $V$ and $\mathbf{z}$ is the enumeration of all the variables in $V-X$.
    \item If $y\notin X$, we set $tr(D_Xy):= \forall\mathbf{z}\forall\mathbf{z'}((A\mathbf{v}\wedge A\mathbf{v}[\mathbf{z'}/\mathbf{z}])\to y = y')$, where $\mathbf{v},\mathbf{z}$ are as in part (d), $\mathbf{z'}$ and $y'$ are the corresponding fresh $V'$-copies of $\mathbf{z}$ and $y$ respectively.\footnote{Furthermore, $A\mathbf{v}[\mathbf{z'}/\mathbf{z}]$ denotes the formula that is obtained by replacing the variables $\mathbf{z}$ by $\mathbf{z'}$ in the formula $A\mathbf{v}$}. If $y\in X$, we set $tr(D_Xy):=\bigwedge_{x\in X}x=x$ (or any other tautology with free variables $X$).
\end{itemize}
The last clause for $y\in X$ ensures that the free variables of any first-order formula of the form $tr(\varphi)$ are contained in $V$; the copied variables $V'$ are only used in quantification and never occur freely in such formulas. There is a one-to-one correspondence between dependence models for $(V,\tau)$ and structures in the language $\tau\cup\{A\}$:if $\mathbb{M}=(M,A)$ is a dependence model then the expansion $T(\mathbb{M})$ of $M$ with the interpretation $I(A):=\{s(\mathbf{v})\;|\;s\in A\}$ is a $\tau\cup\{A\}$-structure, Where on the first-order structure $T(\mathbb{M})$, assignments are functions $\alpha:\mathrm{Var}\to|M|$ for some countably infinite set of variables $\mathrm{Var}$ containing $V\cup V'$. Conversely, for every $\tau\cup\{A\}$ structure $M$, its $\tau$-reduct together with the team $A=\{s:V\to M\;|\;s(\mathbf{v})\in I(A)\}$ forms the unique dependence model $\mathbb{M}$ such that $T(\mathbb{M})=M$. It follows that $\mathbb{M},s\models\varphi$ iff $T(\mathbb{M}),s^+\models tr(\varphi)$, where $s^+$ is any assignment $\mathrm{Var}\to M$ extending $s$. This translation easily adapts to the other local dependence logics discussed in this paper by setting $tr(x=y):=\;x=y$ and $tr(\mathbf{x}\in\mathbf{y}):=\; \exists\mathbf{v'}(A\mathbf{v'}\wedge\bigwedge_{i\leq|\mathbf{x}|}x_i=y'_i)$.
\end{definition}

\section{\textbf{Characterization}}
In \cite{BaltagvBenthemLFD}, it was stated as an open problem to find a bisimulation-invariance theorem characterizing precisely which formulas in $FOL[V\cup V',\tau\cup\{A\}]$ are equivalent to the $tr$-translation of an LFD-formula over standard structures.\footnote{There is also a \textit{modal translation} of LFD into FOL that extends the well-known standard translation of modal logic into the 2-variable fragment of FOL. A similar characterization theorem can be proved via this translation and the relational semantics for LFD, as our notion of bisimulation as well as the one proposed in \cite{local_deps} are naturally formulated on dependence models as well as their modal counterparts.} The following notion of \textit{dependence bisimulation} exactly characterizes $LFD$ as the largest fragment of $FOL$ invariant under this notion. Say that a set of variables $X$ is dependence-closed at $s'$ if $D^{s'}_X:=\{y\in V\;|\;;s'\models D_Xy\}=X$, or equivalently if $s'\models D_Xy$ implies $y\in X$.

\begin{definition}{(\textbf{Dependence Bisimulation})} Let $\mathbb{M},\mathbb{M'}$ be dependence models.\footnote{Clearly this notion also makes sense for associated structures of the form $T(\mathbb{M})$, as well as standard relational models.} We say that a non-empty relation $Z\subseteq A\times A'$ is a \textit{dependence-bisimulation} if for every $(s,s')\in Z$ the following conditions hold:
\begin{itemize}
    \item [(\textbf{AH})] $s\models P\mathbf{x}$ iff $s'\models P\mathbf{x}$
    \item [(\textbf{Forth})] For every $t\in A$, the set $\{v\in V\;|s=_vt\}$ is dependence-closed at $s'$, and\newline
    there is some $t'\in A$ such that
    \begin{enumerate}
            \item [(1)] $(t,t')\in Z$
            \item [(2)] For all $v\in V$, $s=_vt$ implies $s'=_vt'$
        \end{enumerate}
    \item [(\textbf{Back})] symmetric to the (Forth) clause
\end{itemize}
\end{definition}

Note that dependence bisimulations are always \textit{total} bisimulations, meaning that every state is related to some other by the bisimulation, in either model.ddd

\begin{proposition} LFD-formulas are invariant under dependence bisimulations.
\begin{proof}
Let $\mathbb{M},\mathbb{M'}$ be dependence models and $Z\subseteq A\times A'$ a dependence bisimulation with $(s,s')\in Z$ and $\varphi\in$ LFD. We show that $s\models\varphi$ iff $s'\models\varphi$ by induction on the complexity of $\varphi$; the atomic case is immediate by the (AH)-clause and the Boolean cases are also trivial. For the other cases, We show only one direction as the other direction is symmetric.
\begin{itemize}
    \item [($\mathbb{D}_X\psi$)] Suppose that $s\models\mathbb{D}_X\psi$ and let $s'=_Xt'$ for some $t'\in A'$. By the (Back)-clause there is some $t\in A$ such that $s=_Xt$ and $(t,t')\in Z$. It follows that $t\models\psi$ and hence $t'\models\psi$ by $(IH)$.
    \item [($D_Xy$)] Suppose that $s\models D_Xy$ and let $s'=_Xt'$ for some $t'\in A'$. We want to show that $s'=_yt'$. By the (Back)-clause there is some $t\in A$ satisfying (1)-(3). In particular $(t,t')\in Z$ and $s=_Zt$ for $Z=\{v\in V\;|\;s'=_vt'\}$, where $X\subseteq Z$ by assumption. So by monotonicity of dependence, we have $s\models D_Zy$ and by (iii) $Z$ must be dependence-closed at $s$. Hence $y\in Z$, i.e. $s'=_yt'$
\end{itemize}
\end{proof}
\end{proposition} 
Another notion of bisimulation for LFD along more standard lines has been proposed by Gr\"adel and P\"utzst\"uck in \cite{local_deps} that subsumes dependence \textit{atoms} under the atomic harmony (AH) clause, in effect building invariance of dependence atoms into the definition. That is, their definition is exactly the same as definition 2.1. except that instead of the dependence-closed condition they have an extra condition "$s\models D_Xy$ iff $s'\models D_Xy$". This is in a way more standard, as local dependence formulas (from a modal perspective) behave like atoms. Yet \textit{semantically}, local dependence formulas do not behave like atoms because their semantics is not specified by an interpretation-function from outside. As it turns out, the two notions of bisimulation are equivalent, but differ in the corresponding procedures for bisimilarity checking. By the above invariance result (proposition 2,1), dependence bisimulations are also bisimulations in their sense. Conversely, the equivalence "$s\models D_Xy$ iff $s'\models D_Xy$" clearly implies the dependence-closedness condition in both directions, showing that these notions of bisimulation are equivalent. It follows that the proof given in \cite{local_deps} also shows that LFD is the dependence bisimulation-invariant fragment of FOL.

\begin{theorem}{\textbf{Van Benthem Characterization}}\\ $LFD[V,\tau]$ is the largest fragment of $FOL[V\cup V',\tau\cup\{A\}]$ that is invariant under dependence bisimulations.

\end{theorem}
The difference comes out in the corresponding procedures for \textit{bisimilarity-checking}. To check whether a pair $(s,s')$ satisfies the condition that "$s\models D_Xy$ iff $s'\models D_Xy$", one has to compute \textit{for each subset} $X\subseteq V$ the dependence-closures $D^s_X=\{y\in V\;|\;s\models D_Xy\},D^{s'}_X$ (by model checking the respective models, see \cite{local_deps} for an algorithm) and check whether they are equal $D^s_X=D^{s'}_X$. On the other hand, to check whether the dependence-closed condition holds in both ways, we only have to compute dependence-closures $D^s_X, D^{s'}_Y$ for maximal sets of the form $X=\{v\in V\;|\;s=_vt\}$ (or $Y=\{v\in V\;|\;s'=_vt'\}$) and check whether $D^s_X=X$ (or $D^{s'}_Y=Y$). Clearly, the collection of maximal variable sets of the form $X$ or $Y$ is in general a proper subset of the full powerset $\mathcal{P}(V)$. Hence less model-checking is needed to check for the existence of a dependence-bisimulation, than for a bisimulation in the sense of \cite{local_deps}.\newline\newline
Dependence bisimulations generalise naturally to extensions of LFD. For instance, we can extend $LFD$ with the equality relation $=$, yielding the logic $LFD^=$, which is shown to be a conservative reduction class and hence undecidable in \cite{local_deps}. We obtain a natural notion of $LFD^+$-bisimulation by simply requiring the (AH) clause from the definition of dependence bisimulations to also range over atomic formulas of the form $x=y$. We can easily extend the above result to a similar characterization of $LFD^=$ as a fragment of FOL. Interestingly, over full dependence models (i.e. standard structures repackaged as dependence models) for a finite set of variables $V$ with $|V|=k$, this notion of $LFD^=$-bisimulation coincides with $k$-potential isomorphism, which is known to characterize the $k$-variable fragment of first-order logic.\newline\newline
Recently, Gr\"adel and P\"utzst\"uck introduced a family of local dependence logics, where the base logic is LFD without dependence atoms, and new logics are obtained by introducing local versions of dependence properties (and their negations). One of these logics is local inclusion-exclusion logic (denoted by $L[\in,\notin]$ in \cite{local_deps}), which extends the above-mentioned base language with local inclusion atoms $\mathbf{x}\in\mathbf{y}$ (exclusion atoms $\mathbf{x}\notin\mathbf{y}$ being their negations) for tuples of variables $\mathbf{x},\mathbf{y}$ of the same length, with the following semantics:
\[(M,A),s\models\mathbf{x}\in\mathbf{y}\qquad\textrm{iff}\qquad\textrm{there is some}\;t\in A\;\textrm{such that}\;s(\mathbf{x})=t(\mathbf{y})\]
Again, the authors give the same kind of bisimulation for this logic, subsuming invariance for inclusion atoms in the atomic harmony clause. As it turns out, also here we can make a similar move as we did with dependence bisimulations.
\begin{definition}{(\textbf{Inclusion Bisimulation})} Let $\mathbb{M},\mathbb{M'}$ be dependence models. A nonempty relation $Z\subseteq A\times A'$ is an \textit{inclusion-bisimulation} if the following conditions hold for all $(s,s)\in A$:
\begin{itemize}
    \item [(\textbf{AH})] $s\models P\mathbf{x}\qquad\textrm{iff}\qquad s'\models P\mathbf{x}$
    \item [(\textbf{Forth})] For every $t\in A$ there is some $t'\in A$ such that
    \begin{enumerate}
        \item [(1)] $(t,t)\in Z$
        \item [(2)] For all $v\in V$, $s=_vt$ implies $s'=_vt'$
        \item [(3)] For all $x,y\in V$ with $x\ne y:\;\; s(x)=t(y)\quad\textrm{implies}\quad s'(x)=t'(y)$
    \end{enumerate}
    \item [(\textbf{Back})] symmetric to (Forth)
\end{itemize}
\end{definition}

\vspace{0.05in}

\begin{proposition} $L[\in,\notin][V,\tau]$ is the largest fragment of $FOL[V\cup V',\tau\cup\{A\}]$ that is invariant under inclusion bisimulation.
\begin{proof}
For invariance, suppose that $s\models\mathbf{x}\in\mathbf{y}$, i.e. $s(\mathbf{x})=t(\mathbf{y})$ for some $t\in A$. By (Forth) there is some matching $t'\in A$ satisfying (1)-(3). Let $i\leq|\mathbf{x}|=|\mathbf{y}|$; if $x_i=y_i$ then by (2) $s'(x_i)=t'(y_i)$ and if $x_i\ne y_i$ then by (3) $s'(x_i)=t'(y_i)$ (since $s(x_i)=t(y_i)$). The other direction is symmetric. By standard model-theoretic arguments, the converse  holds by showing that $L[\in,\notin]$-equivalence implies bisimilarity over $\omega$-saturated $\tau\cup\{A\}$-structures.
\end{proof}
\end{proposition}

\section{\textbf{Finite Model Property}}
We show that LFD has the FMP (w.r.t dependence models). We use a similar proof-strategy as in \cite{grAdel_1999} to prove FMP for GF. Let $\varphi\in LFD$ be satisfiable, and let $\Phi=Cl(\{\varphi\})$. So we consider $LFD$ over the finite vocabulary $(V_{\varphi},\tau)$, where $\tau$ is the collection of predicate symbols occurring in $\varphi$. As $V$ is finite, let $|V_{\varphi}|=k$. Then by our tree model property, there is a dependence model $\mathbb{M}=(M,A)$ satisfying $\varphi$ at the root assignment $v_{\langle\Sigma_0\rangle}$, where $M$ is a $k$-tree of finite branching degree. Moreover, we may assume that there is some type model $\mathfrak{M}$ generating the model $\mathbb{M}$, as in the proof of theorem 1.1. We cut the underlying $k$-tree $M$ to a finite structure, and subsequently use Herwig's theorem to generate out of this a finite dependence model satisfying $\varphi$. Recall that $A=\{v_{\pi}\;|\;\pi\;\textrm{a good path}\}$, where $T=\{\pi\;|\;\pi\;\textrm{a good path}\}$ is the tree (of finite branching degree) of traces the type model $\mathfrak{M}$. We define the cut-off model $\mathbb{M}_{cut}=((O_{cut},I),A_{cut})=(M_{cut},A_{cut})$:
\begin{align*}
    &A_{cut}:=\{v_{\pi}\in A\;|\;lh(\pi)\leq 3\}\\
    &O_{cut}:=\bigcup\{v_{\pi}[V_{\varphi}]\;|\;v_{\pi}\in A_{cut}\}\\
    &I^{M_{cut}}(P):=I^{M}(P)\restriction O_{cut}^{ar(P)}
\end{align*}
Now $\mathbb{M}_{cut}$ is \textit{finite} because our $k$-tree $M$ is of finite branching degree. The truth lemma no longer holds of the cut-off model $\mathbb{M}_{cut}$. However, it does still satisfy the following \textit{restricted truth lemma}:

\begin{lemma} For all $v_{\pi}\in A_{cut}$ of $lh(\pi)\leq 2:\;D_Xy\in last(\pi)$ iff $\mathbb{M}_{cut},v_{\pi}\models D_Xy$

\begin{proof}
($\rightarrow$) Suppose that $D_Xy\in last(\pi)$. By the truth lemma, $\mathbb{M},v_{\pi}\models D_Xy$. As $A_{cut}\subseteq A$, it follows from the semantics that $\mathbb{M}_{cut},v_{\pi}\models D_Xy$. ($\leftarrow$) By contraposition. So suppose that $D_Xy\not\in last(\pi)$. Define the sequence $\pi^+:=(\pi,X,last(\pi))$; this is a good path as $last(\pi)\sim_X last(\pi)$ is always true. Observe that $lh(\pi^+)=lh(\pi)+1\leq 2+1=3$, and hence $v_{\pi^+}\in A_{cut}$. By the recursive definition of $O$ and $A$, we see that $v_{\pi}=_Xv_{\pi^+}$ and $v_{\pi}\ne_yv_{\pi^+}$ as $X\subseteq D^{last(\pi)}_X$ and $y\not\in D^{last(\pi)}_X$ respectively. By the semantics, this means that $\mathbb{M}_{cut},v_{\pi}\not\models D_Xy$.
\end{proof}
\end{lemma}

Before turning to Herwig's theorem, we expand the cut-off model with interpretations for new relation symbols $R^{X,y}$ for $X\cup\{y\}\subseteq V_{\varphi}$ that we use to encode the dependence atoms $D_Xy$. Let $\tau^+$ be the smallest language containing $\tau$ and an $|X|$-ary relation $R^{X,y}$ for each $X\cup\{y\}\subseteq V_{\varphi}$. Observe that $(V_{\varphi},\tau^+)$ is still a finite vocabulary. We expand the underlying $\tau$-structure $M_{cut}$ of the cut-off model to a $\tau^+$ structure by setting  $I^{M_{cut}}(R^{X,y}):=\{v_{\pi}(\mathbf{x})\;|\;D_Xy\in last(\pi)\}$, whence the restricted truth lemma becomes expressible as a big conjunction of formulas of the form $R^{X,y}\mathbf{x}\leftrightarrow D_Xy$ in the $LFD[V_{\varphi},\tau^+]$.\newline\newline
Herwig's theorem \cite{Herwig1998ExtendingPI} on extending partial isomorphism is a result in the finite model theory of first-order relational languages. It tells us that any finite structure with some set of partial isomorphisms on it has a finite extension in which all these partial isomorphisms extend to automorphisms of the extension. This theorem has already been used to show the FMP of GF \cite{grAdel_1999}. Herwig's theorem is formulated in terms of first-order relational languages, and so we will apply it not to the dependence model $\mathbb{M}_{cut}$ but to the $\tau^+\cup\{A\}$-structure $T(\mathbb{M}_{cut})$.

\begin{theorem}{\textbf{Herwig}}\\
Let $\sigma$ be a finite relational language, $C$ a finite $\sigma$-structure and $\{p_1,...,p_k\}$ a (finite) set of partial isomorphisms on $C$. Then there exists a \textit{finite} extension $C^+$ of $C$ that satisfies the following conditions:
\begin{itemize}
    \item [(i)] Every $p_i$ extends to a unique automorphism $\widehat{p_i}$ of $C^+$. This yields a subgroup $\langle\widehat{p_1},...,\widehat{p_k}\rangle$ of the automorphism group of $C^+$.
    \item [(ii)] If a tuple $\mathbf{a}=(a_1,....,a_r)$ from $C^+$ is live or $r=1$, then there exists an automorphism $f\in\langle\widehat{p_1},...,\widehat{p_k}\rangle$ such that for each $i\leq r$, $f(a_i)\in C$.
    \item [(iii)] If $\exists f\in\langle\widehat{p_1},...,\widehat{p_k}\rangle$ and $a,b\in C$ such that $f(a)=b$, then either $f=id$ or there is a unique $p\in\langle p_1,...,p_k\rangle$ such that $\widehat{p}=f$ and $p(a)=b$.
\end{itemize}
\end{theorem}
where $\langle p_1,...,p_k\rangle$ is the collection of all partial isomorphisms that can be obtained by composing the $p_i$ with their inverses (or rather converses, because the $p_i$ are functional relations). Note that $\langle p_1,...,p_k\rangle$ is \textit{not} a group as it does not include the identity on $C$, nor is the case that $p\circ p^{-1}$ is the identity on $C$.\newline\newline
Condition (iii) is in need of further clarification. In words, it says that elements in the submodel $C$ are only mapped to each other by some $f\in\langle f_1,...,f_n\rangle$ if this is forced given the choice of partial isomorphisms. Uniqueness of $p$ is ensured by the fact that the map $\widehat{(\;)}$ extends to a bijective map $\widehat{(\;)}:\langle p_1,...,p_k\rangle\to\langle\widehat{p_1},...,\widehat{p_k}\rangle$ that commutes with the operations $\circ,(\;)^{-1}$ (and the identity $id$). By condition (i), $\widehat{(\;)}$ is defined on the subset $\{p_1,...,p_k\}$. Set $\widehat{p^{-1}}:=\widehat{p}^{-1}$ and $\widehat{p\circ p'}:=\widehat{p}\circ\widehat{p'}$; so commutation follows by definition. It immediately follows that the map is injective. For surjectivity, let $f\in\langle\widehat{p_1},...,\widehat{p_k}\rangle$. By definition $f=\widehat{p_{i_1}}^{\epsilon_1}\circ...\circ \widehat{p_{i_m}}^{\epsilon_m}$ for some $\{i_1,...,i_m\}\subseteq\{1,...,k\}$ and $\epsilon_j\in\{-1,1\}$ for each $j\leq m$. Define $p:=p_{i_1}^{\epsilon_1}\circ...\circ p_{i_m}^{\epsilon_m}\in\langle p_1,...,p_k\rangle$. Now observe:
\[\widehat{p}=\reallywidehat{p_{i_1}^{\epsilon_1}\circ...\circ p_{i_m}^{\epsilon_m}}=\widehat{p_{i_1}^{\epsilon_1}}\circ...\circ\widehat{p_{i_m}^{\epsilon_m}}=\widehat{p_{i_1}}^{\epsilon_1}\circ...\circ\widehat{p_{i_m}}^{\epsilon_m}=f\]
We proceed with specifying a choice of partial isomorphisms on the cut-off model. If $\pi$ is a good path of $lh(\pi)=3$ and $last(\pi)=\Delta$, then there is a \textit{unique} partial isomorphism $p_{\pi}:v_{\pi}[V_{\varphi}]\to v_{\pi_{\Delta}}[V_{\varphi}]$ such that $p_{\pi}\circ v_{\pi}=v_{\pi_{\Delta}}$, where $\pi_{\Delta}:=\langle\Sigma_0,\emptyset,\Delta\rangle$ so $lh(\pi_{\Delta})=2$. We pick the finite set of partial isomorphisms $\{p_{\pi}\;|\;\pi\;\textrm{good path of}\;lh(\pi)=3\}=\{p_1,...,p_k\}$. The following proposition tells us what kind of partial isomorphisms 
are in $\langle p_1,...,p_k\rangle$.

\vspace{0.1in}

\begin{lemma} If $p\in\langle p_1,...,p_k\rangle$ with $pv_{\pi}=_Xv_{\pi'}$, then there are $v_{\rho},v_{\rho'}\in A_{cut}$ such that $pv_{\rho}=v_{\rho'}$ with $last(\rho)=last(\rho')$, and $v_{\rho}=_Xv_{\pi}$, $v_{\rho'}=_Xv_{\pi'}$.
\begin{proof}

Recall that $dom(v_{\pi_0})=V_{\varphi}$ for all good paths $\pi_0$. Let $p\in\langle p_1,...,p_k\rangle$ such that $pv_{\pi}=_Xv_{\pi'}$. By definition $p=p_{i_m}^{\epsilon_m}\circ...\circ p_{i_1}^{\epsilon_1}$ for some $\{i_1,...,i_m\}\subseteq\{1,...,k\}$ and $\epsilon_j\in\{-1,1\}$ for each $1\leq j\leq m$. Note that for each $j\leq m$ we have that $p_{i_j}\in\{p_1,...,p_k\}=\{p_{\pi}\;|\;\pi\;\textrm{a good path of}\;lh(\pi)=3\}$, so $p_{i_j}^{\epsilon_j}\circ v_{\pi_{j-1}}=v_{\pi_j}$ for some $v_{\pi_{j-1}},v_{\pi_j}\in A_{cut}$ (so $dom(p_{i_j}^{\epsilon_j})=v_{\pi_{j-1}}[V_{\varphi}]$ and $cod(p_{i_j}^{\epsilon_j})=v_{\pi_j}[V_{\varphi}]$) with $last(\pi_{j-1})=last(\pi_j)$. In particular, there are $v_{\pi_0},v_{\pi_1}\in A_{cut}$ such that $last(\pi_0)=last(\pi_1)$ and $p_{i_1}^{\epsilon_1}v_{\pi_0}=v_{\pi_1}$. It follows that $v_{\pi}=_Xv_{\pi_0}$ and so $pv_{\pi_0}=_Xv_{\pi'}$, i.e.
\[pv_{\pi_0}=p_{i_m}^{\epsilon_m}\circ...\circ p_{i_1}^{\epsilon_1}v_{\pi_0}=p_{i_m}^{\epsilon_m}\circ...\circ p_{i_2}^{\epsilon_2}v_{\pi_1}=_Xv_{\pi'}\]
This was the base case for an inductive argument up to $m$. So let $j\leq m$ and suppose that $v_{\pi_j}\in A_{cut}$ with $last(\pi_j)=\Delta$ and 
\[pv_{\pi_0}=p_{i_1}^{\epsilon_1}\circ ... \circ p_{i_{j+1}}^{\epsilon_{j+1}}v_{\pi_j}=_Xv_{\pi'}\]
Now recall that $p_{i_{j+1}}^{\epsilon_{j+1}}v_{\pi_{j+1}}=v_{\pi_{j+2}}$ for some $v_{\pi_{j+2}}\in A_{cut}$ with $last(\pi_{j+2})=last(\pi_{j+1})=\Delta$. Moreover, it follows that $p_{i_1}^{\epsilon_1}\circ...\circ p_{i_{j+2}}^{\epsilon_{j+2}}v_{\pi_{j+1}}=_Xv_{\pi'}$. Hence by induction there is some $v_{\pi_m}\in A_{cut}$ such that $last(\pi_m)=last(\pi_0)=\Delta$ and 
\[v_{\pi_m}=p_{i_m}^{\epsilon_m}\circ...\circ p_{i_1}^{\epsilon_1}v_{\pi_0}=pv_{\pi_0}=_Xv_{\pi'}\]
then for $\rho=\pi_0$ and $\rho'=\pi_m$ we have proved the lemma.
\end{proof}
\end{lemma}

The associated first-order structure $T(\mathbb{M}_{cut})$ of the Herwig extension is a finite model in a finite relational language $\tau^+\cup\{A\}$, and $\{p_1,...,p_k\}$ is a finite set of partial isomorphisms on it. Hence, by Herwig's theorem, there exists a \textit{finite} extension $T(\mathbb{M}_{cut})^+$ of this structure, the \textit{Herwig extension}, satisfying conditions (i)-(iii) w.r.t $\{p_1,...,p_k\}$. It is easy to see that the Herwig extension corresponds in the canonical way to a dependence model; let $\mathbb{M}_{cut}^+:=(M_{cut}^+,A_{cut}^+)$, where $M_{cut}^+$ is the $\tau^+$ reduct of $T(\mathbb{M}_{cut})^+$, and where $A_{cut}^+:=\{s:V_{\varphi}\to M_{cut}^+\;|\;s(\mathbf{v})\in I(A)\}$. It follows that $T(\mathbb{M}_{cut}^+)=T(\mathbb{M}_{cut})^+$. We can show that every assignment $A_{cut}^+$ can be automorphically mapped to some path assignment $v_{\pi}$ where $lh(\pi)\leq 2$. We will refer to this fact as the \textit{level 2 lemma}: 

\vspace{0.1in}

\begin{lemma} For every $s\in A_{cut}^+$ there is an $f\in\langle\widehat{p_1},...,\widehat{p_k}\rangle$ such that $f\circ s=v_{\pi}\in A_{cut}$ where $lh(\pi)\leq 2$.
\begin{proof}
Let $s\in A_{cut}^+$. Then the tuple $s(\mathbf{v})\in I(A)$ is live in $T(\mathbb{M}_{cut}^+)$. Hence by condition (ii) there is some $f\in\langle\widehat{p_1},...,\widehat{p_k}\rangle$ such that $fs(\mathbf{v})$ is a tuple of objects from the old domain $O_{cut}$ with $fs(\mathbf{v})\in I(A)$, as $f$ is an isomorphism on $T(\mathbb{M}_{cut}^+)$. It follows that $fs(\mathbf{v})=v_{\pi}(\mathbf{v})$ for some $v_{\pi}\in A_{cut}$. But if $lh(\pi)=3$ and $last(\pi)=\Delta$, then by (i) there is an automorphism $\widehat{p_{\pi}}$ such that $\widehat{p_{\pi}}f\in\langle\widehat{p_1},...,\widehat{p_k}\rangle$ and $\widehat{p_{\pi}}fs=v_{\pi_{\Delta}}$, where $lh(\pi_{\Delta})=2$. Hence $\exists g\in\langle\widehat{p_1},...,\widehat{p_k}\rangle$ such that $gs=v_{\pi}$ for some $v_{\pi}\in A_{cut}$ of $lh(\pi)\leq 2$.\footnote{If $dom(s)=V_{\varphi}$ then it makes sense to say that $g\circ s=v_{\pi}$ as $dom(v_{\pi})=V_{\varphi}$ as well. However, if $s$ is an assignment in the wider sense that $dom(s)\supset V_{\varphi}$ then what we mean by writing $g\circ s=v_{\pi}$ is that $g\circ s=_{V_{\varphi}}v_{\pi}$. We will sometimes skip the functional composition sign $\circ$ for readability, and e.g. write $gs$ for the function $g\circ s$.}
\end{proof}
\end{lemma}

We want to generalise the notion of 'underlying type' to all assignments in the Herwig extension, to be able to prove a restricted truth lemma for dependence atoms for all assignments in $A_{cut}^+$. Set $type(s):=last(\pi)$ for  $s\in A_{cut}$ (i.e. $s$ is of the form $v_{\pi}$) and $type(s):=last(\pi)$, where $v_{\pi}\in A_{cut}$ such that $\exists f\in\langle\widehat{p_1},...,\widehat{p_k}\rangle$ with $f\circ s=v_{\pi}$ (which exists by the level 2 lemma above). This is well-defined, for let $g\in\langle\widehat{p_1},...,\widehat{p_k}\rangle$ be some other automorphism with $gt=v_{\pi'}\in A_{cut}$. Then $f\circ g^{-1}$ is an automorphism in the subgroup $\langle\widehat{p_1},...,\widehat{p_k}\rangle$ that maps elements in the old domain $O_{cut}$ to each other as $fg^{-1}\circ v_{\pi'}=v_{\pi}$, hence by (iii) there must be a \textit{unique} $p\in\langle p_1,...,p_k\rangle$ such that $\widehat{p}=fg^{-1}$ and $pv_{\pi'}=v_{\pi}$. Lemma 3.2 then says that there must be $v_{\rho},v_{\rho'}\in A_{cut}$ with $v_{\pi}=_{V_{\varphi}}v_{\rho}$, $v_{\pi'}=_{V_{\varphi}}v_{\rho'}$ and $last(\rho)=last(\rho')$. By the truth lemma, it follows that $last(\pi)=last(\rho)=last(\rho')=last(\pi')$. Observe that this definition of $type(\;)$ entails that $s(\mathbf{x})\in I(R^{X,y})$ iff $D_Xy\in type(t)$. We know show that the concrete relations $=_X$ on assignments works well with the abstract relations $\sim_X$ on their underlying types.

\begin{lemma} If $s,t\in A_{cut}^+$ with $s=_Xt$, then $type(s)\sim_X type(t)$.
\begin{proof}
Let $s,t\in A_{cut}^+$ with $s=_Xt$. By the level 2 lemma, there is $f\in\langle\widehat{p_1},...,\widehat{p_k}\rangle$ such that $fs=v_{\pi}\in A_{cut}$ with $lh(\pi)\leq 2$, so $type(s)=last(\pi)$. As $f$ is an isomorphism, $ft\in A_{cut}^+$ as well with $fs=v_{\pi}=_Xft$. By applying the level 2 lemma to $ft$, there is $g\in\langle\widehat{p_1},...,\widehat{p_k}\rangle$ such that $gft=v_{\pi'}\in A_{cut}$ with $lh(\pi')\leq 2$, so $type(t)=last(\pi')$. Moreover, we know that $gv_{\pi}$ is an assignment (not necessarily in $A_{cut}$) such that $gv_{\pi}=gfs=_Xgft=v_{\pi'}$, since $fs=_Xft$ and $g$ is an isomorphism. Hence $g$ maps elements in the old domain $O_{cut}$ to each other so by condition (iii) there must be a unique $p\in\langle p_1,...,p_k\rangle$ such that $\widehat{p}=g$ and $pv_{\pi}=_Xv_{\pi'}$. By lemma 2.2 there are $v_{\rho},v_{\rho'}\in A_{cut}$ such that $v_{\pi}=_Xv_{\rho}$, $v_{\pi'}=_Xv_{\rho'}$ and $last(\rho)=last(\rho')$. It follows that $type(s)=last(\pi)\sim_X last(\rho)=last(\rho')\sim_X last(\pi')=type(t)$
\end{proof}
\end{lemma}

\begin{lemma} For all $s\in A_{cut}^+$:$\;D_Xy\in type(s)$ iff $\mathbb{M}_{cut}^+,s\models D_Xy$. 

\begin{proof}
($\leftarrow$) We prove the equivalent statement that $\mathbb{M}_{cut}^+,s\models\bigwedge_{R^{X,y}\in\tau^+} R^{X,y}\mathbf{x}\leftrightarrow D_Xy$. By the level 2 lemma, there is $f\in\langle\widehat{p_1},...,\widehat{p_k}\rangle$ such that $fs=v_{\pi}\in A_{cut}$ with $lh(\pi)\leq 2$. By the restricted truth lemma and the first-order translation, we obtain that $T(\mathbb{M}_{cut}),v_{\pi}\models tr(\neg R^{X,y}\mathbf{x}\to\neg D_Xy)$. But observe that 
\begin{align*}
    tr(\neg R^{X,y}\mathbf{x}\to\neg D_Xy)\;=\;\neg R^{X,y}\mathbf{x}\to tr(\neg D_Xy)\;\equiv\;& R^{X,y}\mathbf{x}\vee\exists\mathbf{z},\mathbf{z'}(A\mathbf{v}\wedge A\mathbf{v'}[\mathbf{z}/\mathbf{z'}]\wedge y\ne y')\\
    \;\equiv\;&\exists\mathbf{z},\mathbf{z'}(R^{X,y}\mathbf{x}\vee(A\mathbf{v}\wedge A\mathbf{v'}[\mathbf{z}/\mathbf{z'}]\wedge y\ne y'))
\end{align*}
is an \textit{existential} first-order formula. Hence by the dualized version of the Łoś-Tarski theorem, this still holds in the Herwig \textit{extension}, i.e. $T(\mathbb{M}_{cut}^+),v_{\pi}\models\neg R^{X,y}\mathbf{x}\to tr(\neg D_Xy)$. As $f$ is an isomorphism on $T(\mathbb{M}_{cut}^+)$ and $fs=v_{\pi}$, we get that $T(\mathbb{M}_{cut}^+),s\models\neg R^{X,y}\mathbf{x}\to tr(\neg D_Xy)$, as desired.\newline\newline
($\to$) Suppose for contradiction that $s\models R^{X,y}\mathbf{x}\wedge \neg D_Xy$. Then $D_Xy\in type(s)$ and there is some $t\in A_{cut}^+$ such that $s=_Xt$ and $s\ne_yt$. Lemma 3.4 tells us that $type(s)\sim_X type(t)$ and so $D_Xy\in type(t)$ as well. Let $Y:=\{v\in V_{\varphi}\;|\;s=_vt\}$ be the maximal set of variables on which $s$ and $t$ agree. By the level 2 lemma, there is an automorphism $f\in\langle\widehat{p_1},...,\widehat{p_k}\rangle$ such that $fs=v_{\pi}\in A_{cut}$ and $ft\in A_{cut}^+$. Applying the same lemma to the assignment $ft$, we get an automorphism $g\in\langle\widehat{p_1},...,\widehat{p_k}\rangle$ such that $gft=v_{\pi'}\in A_{cut}$ and $gfs\in A_{cut}^+$. As both $f,g$ are bijective we get that
\[s=_v t\qquad\textrm{iff}\qquad (v_{\pi}=)\;fs=_vft\qquad\textrm{iff}\qquad gv_{\pi}=_vgft\;(=v_{\pi'})\]
which means that $Y=\{v\in V_{\varphi}\;|\;v_{\pi}=_vft\}=\{v\in V_{\varphi}\;|\;gv_{\pi}=_vv_{\pi'}\}$ is also the maximal set of variables on which $v_{\pi}$ and $ft$, respectively $gv_{\pi}$ and $v_{\pi'}$ agree. In particular, it follows that (a) $ft\ne_yv_{\pi}=fs$ because $t\ne_y s$. Note that $v_{\pi}(x)=ft(x)$ for all $x\in X$ and so $v_{\pi}(x)\overset{g}{\mapsto}v_{\pi'}(x)$ maps elements in $O_{cut}$ to each other for each $x\in X$. By condition (iii) then, there must be a \textit{unique} $p\in\langle p_1,...,p_k\rangle$ such that $\widehat{p}=g$ and $pv_{\pi}=_Xv_{\pi'}$. Using Lemma 2.2, we know there must be $v_{\rho},v_{\rho'}\in A_{cut}$ such that $v_{\pi}=_Xv_{\rho}$, $v_{\pi'}=_Xv_{\rho'}$, $pv_{\rho}=v_{\rho'}$ and $last(\rho)=last(\rho')$. Now $D_Xy\in type(s)\cap type(t)=last(\pi)\cap last(\pi')$ so by clause (2) of the recursive definition of path assignments on page 4, we obtain that $v_{\rho'}=_yv_{\pi'}$ and (b) $v_{\pi}=_yv_{\rho}$. The former equality in turn implies that
\begin{equation*}{(c)}
    \qquad v_{\rho}=p^{-1}v_{\pi_m}=\widehat{p^{-1}}v_{\pi_m}=_y\widehat{p^{-1}}v_{\pi'}=\widehat{p^{-1}}gft=g^{-1}gft=ft
\end{equation*}
as $g^{-1}=\widehat{p}^{-1}=\widehat{p^{-1}}\in\langle\widehat{p_1},...,\widehat{p_k}\rangle$ is bijective. In conclusion, concatenating (a), (b) and (c), we have derived the contradiction $ft\ne_yv_{\pi}=_yv_{\rho}=_yft$.
\end{proof}
\end{lemma}

\begin{proposition} The dependence models $\mathbb{M}$ and $\mathbb{M}_{cut}^+$ are dependence-bisimilar.
\begin{proof}
We show that the relation $Z\subseteq A_{cut}^+\times A$ defined by $Z:=\{(s,v_{\pi})\;|\;type(s)=last(\pi)\}$ is an LFD-bisimulation in the sense of $\cite{local_deps}$ and hence, by our remark above, also a dependence bisimulation. Pick an arbitrary pair $(s,v_{\pi})\in Z$. We show that the pair $(s,v_{\pi})$ satisfies atomic harmony (AH), including dependence atoms, and is closed under the (Back) \& (Forth) clauses without the dependence-closedness condition. By the level 2 lemma, $(s,v_{\pi})\in Z$ means that there is some $f\in\langle\widehat{p_1},...,\widehat{p_k}\rangle$ such that $fs=v_{\pi'}\in A_{cut}$ with $lh(\pi')\leq 2$ and $last(\pi)=last(\pi')$. For $\textbf{(AH)}$, observe that the chain of equivalences:
\[s\models_{\mathbb{M}_{cut}^+} P\mathbf{x}\quad \textrm{iff}\quad v_{\pi'}\models_{\mathbb{M}_{cut}^+}P\mathbf{x}\quad \textrm{iff}\quad P\mathbf{x}\in last(\pi')=last(\pi)\quad \textrm{iff}\quad v_{\pi}\models_{\mathbb{M}} P\mathbf{x}\]
holds by the fact that $f$ is an isomorphism with $fs=v_{\pi'}$ and the way we have specified the interpretation $I(P)$ on both models. Let $D^s=\{D_Xy\;|\;s\models D_Xy\}$ and $D^{\Delta}=\{D_Xy\;|\;D_Xy\in\Delta\}$. We show that $D^s=D^{v_{\pi}}$, i.e. "$s\models_{\mathbb{M}_{cut}^+} D_Xy$ iff $v_{\pi}\models_{\mathbb{M}} D_Xy$". Since $f$ is an isomorphism and $s(\mathbf{x})\in I(R^{X,y})$ iff $D_Xy\in type(s)$, we get that $D^{type(s)}=D^{last(\pi')}$. Moreover, we know that $last(\pi)=last(\pi')$, so $D^s=D^{last(\pi)}$. By lemma 3.5 (a restricted truth lemma for the Herwig extension) we can identify the dependence-order of the assignment with the dependence order of the underlying type $D^s=D^{type(s)}$, and similarly $D^{last(\pi)}=D^{v_{\pi}}$ follows from the truth lemma. Hence, we may conclude that $D^s=D^{v_{\pi}}$.\newline\newline
\textbf{(Forth)} Let $t\in A_{cut}^+$ be some assignment in the Herwig extension, and consider $Y=\{v\in V_{\varphi}\;|\;s=_vt\}$ the maximal set of variables on which $s$ and $t$ agree, so $s=_Yt$. By lemma 3.4, it follows that $type(s)\sim_Y type(t)$. But $last(\pi)=last(\pi')=type(s)$, so $\pi^+:=\langle\pi,Y,type(t)\rangle$ is a good path. It follows that $v_{\pi^+}\in A$, with $v_{\pi}=_Y v_{\pi^+}$ and $(t,v_{\pi^+})\in Z$ as $type(t)=last(\pi^+)$.\newline\newline
\textbf{(Back)} Let $v_{\pi''}\in A$, with $Y=\{v\in V\;|\;v_{\pi}=_vv_{\pi''}\}$ the maximal set of relevant variables such that $v_{\pi}=_Yv_{\pi''}$. By the truth lemma and locality, this entails that $last(\pi)\sim_Ylast(\pi'')$ so as $type(s)=last(\pi')=last(\pi)$ we get a good path $\pi'^+:=\langle\pi',Y,last(\pi'')\rangle$. Recall that we may assume (by the level 2 lemma) that $lh(\pi')\leq 2$, so that $lh(\pi'^+)=lh(\pi')+1\leq 2+1=3$ and thus $v_{\pi'^+}\in A_{cut}$ is in the cut-off model. Clearly $v_{\pi'}=_Yv_{\pi'^+}$ and for $t:=f^{-1}v_{\pi'^+}$, we then get that $s=_Yt$ and $(t,v_{\pi''})\in Z$ since $type(t)=last(\pi'^+)=last(\pi'')$.
\end{proof}
\end{proposition}

\begin{corollary}{\textbf{Finite Model Property}}\\
LFD has the finite model property.
\begin{proof}
Let $\varphi$ be a satisfiable LFD-formula with closure $\Phi$ in the language $(V_{\varphi},\tau)$ with $|V_{\varphi}|=k$. By the tree model property, there is a $k$-tree $M$ and a team $A$ such that $\mathbb{M}=(M,A)$ is a dependence model satisfying $\varphi$ at the root assignment $v_{\langle\Sigma_0\rangle}$ (so $\varphi\in\Sigma_0$). Construct the Herwig extension $\mathbb{M}_{cut}^+=(M_{cut},A_{cut})$ as above. By proposition 3.1, the infinite tree $\mathbb{M}$ and the Herwig extension are dependence-bisimilar. As dependence bisimulations are always total, there is some assignment $s\in A_{cut}^+$ that is related by this dependence bisimulation to the root assignment $v_{\langle\Sigma_0\rangle}\in A$. By our invariance result above (proposition 2.1), it follows that $\mathbb{M}_{cut}^+$ is a finite model of $\varphi$.
\end{proof}
\end{corollary}

\section{Conclusion}
We have presented the notion of dependence bisimulation and show that it characterizes the logic LFD as a fragment of FOL. Moreover, we have shown that it is equivalent to another notion of bisimulation LFD proposed in \cite{local_deps}, but more efficient for bisimilarity-checking. We did the same for local inclusion-exclusion logic, a related logic introduced in \cite{local_deps}. Furthermore, we have shown that LFD has the finite model property. We used a tree model property established in \cite{BaltagvBenthemLFD}, and shown how to use Herwig's theorem to a finite part of this tree model to obtain a finite model bisimulating the full infinite tree. In our proof, we have made novel use of Herwig's theorem, which uses more conditions guaranteed by the theorem than its application in the proof of FMP for GF \cite{grAdel_1999}.

\appendix
\section{Modal Finite Model Property }
Dependence models have natural modal counterparts, over which LFD becomes a modal logic. We give a direct proof of the finite model property for LFD w.r.t the modal semantics, using a more general version of Herwig's theorem which additionally guarantees that the Herwig extension omits packed homomorphic images. 

\begin{definition}{(\textbf{Standard Relational Models})} A \textit{standard relational model} is a structure $\mathbb{A}:=(A,\sim_X,D_Xy,P\mathbf{x})$, where for each finite subset $X$ of $V$, $\sim_X$ is a binary relation on $A$, and $D_Xy,P\mathbf{x}$ are unary predicates on $A$ for every $X\cup\{y\}\subseteq V$ and $\mathbf{x}\in V^{ar(P)}$, respectively. This induces relations $D^s:=\{(X,y)\;|\;D_Xy(s)\}$ and $P^s:=\{\mathbf{x}\in V^{ar(P)}\;|\;P\mathbf{x}(s)\}$.\footnote{Alternatively, one can present the modal semantics algebraically, specifying conditions on the valuation. Here, we stick to the presentation of relational models as first-order structures.} These have to satisfy the following conditions:
\begin{enumerate}
    \item [(1)] all relations $\sim_X$ are equivalence relations on $A$
    \item [(2)] all relations $D^s$ satisfy Transitivity and Projection
    \item [(3)] if $s\sim_X t$ and $D_Xy(s)$, then $s\sim_y t$ and $D_Xy(t)$
    \item [(4)] if $s\sim_X t$ and $P\mathbf{y}(s)$ for some $\{y_1, ... ,y_m\}\subseteq X$, then $P\mathbf{y}(t)$
    \item [(5)] $\sim_{\emptyset}$ is the universal relation    \item [(6)] if $s\sim_X t$ and $s\sim_Y t$, then $s\sim_{X\cup Y} t$.
    \item [(7)] if $s\sim_X t$ implies $s\sim_y t$ holds for all $t\in A$, then $D_Xy(s)$ holds.
\end{enumerate}
\end{definition} 

Structures $\mathbb{A}=(A,\sim_X,D_Xy,P\mathbf{x})$ satisfying conditions (1)-(5) are called \textit{(general) relational models}. These form a modal counterpart to the non-standard type model semantics.\footnote{By the standard modal technique of filtration technique tailored to deal with the semantics of the dependence atoms, one can show that every satisfiable LFD-formula has a finite relational model \cite{BaltagvBenthemLFD}.} Note that conditions (3) and (7) together imply that $I(D_Xy):=\{s\in A\;|\;[s]_X\subseteq[s]_y\}$ (where $[s]_X$ is the $\sim_X$ equivalence class of $s$). The semantics are defined almost exactly the same as on dependence models:
\begin{definition}{(\textbf{Semantics})} Truth of a formula $\varphi$ in a (general) relational model $\mathbb{A}=(A,\sim_X,D_Xy,P\mathbf{x})$ at an assignment $s\in A$ is defined by the following clauses  (the Boolean cases are defined as usual):
\begin{align*}
& s\models P\mathbf{x}\;\textrm{iff}\;s\in I^{\mathbb{A}}(P)\\
& s\models\mathbb{D}_X\varphi\;\textrm{iff}\;t\models\varphi\;\textrm{holds for all}\;t\in A \;\textrm{with}\; s \sim_X t\\
& s\models D_Xy\;\textrm{iff}\;s\in I^{\mathbb{A}}(D_Xy)
\end{align*}
where we write $s(\mathbf{x})=(s(x_1),...,s(x_m))$ if $\mathbf{x}=(x_1,...,x_m)$. But note that on standard models, by condition (3) and (7) $s\models D_Xy$ iff $[s]_x\subseteq [s]_y$, where $[s]_X$ is the $\sim_X$ equivalence class of $s$.
\end{definition}

Similar to our $k$-tree model property, by combining proposition A.3 and A.5 from \cite{BaltagvBenthemLFD} one can show that every satisfiable $\varphi$ is satisfiable at the root of a standard relational model $\mathbb{A}=(A,\sim_X,D_Xy,P\mathbf{x})$ such that $A$ is a set of \textit{histories} through some finite general relational model $\mathbb{B}$\footnote{In fact, $\mathbb{B}$ is a finite \textit{filtrated} model, with states $[s]$ being $\equiv_{\Phi}$-equivalence classes ($\Phi$ being some closure set).}, where $A$ forms a \textit{tree} w.r.t. initial segment relation. A history through $\mathbb{B}$ is a sequence $h=(b_0,X_1,...,X_n,b_n)$ (with $b_0\in\mathbb{B}$ fixed) such that $b_0\sim_{X_1}...\sim_{X_n}b_n$ (and each $b_i\in\mathbb{B},X_i\subseteq V)$. Write $last(h)=b_n$ and $lh(h)=n+1$. For each $X\subseteq $ define one-step relations $\sim_X^1$ on $A$ by:
\begin{align*}
    h\sim_X^1h'\qquad\textrm{iff}\qquad &h'=(h,Y,last(h'))\;\textrm{with}\;h\models D_YX\;\textrm{,\;or}\\
    &h=(h',Y,last(h))\;\textrm{with}\;h'\models D_YX
\end{align*}
and consider the structure $\mathbb{A}=(A,\sim_X^1,D_Xy,P\mathbf{x})$. We define the cut-off model as the submodel induced by $A_{cut}\subseteq A$, where as before $A_{cut}:=\{h\in A\;|\;lh(h)\leq 3\}$. Observe that in $(A_{cut},\sim^1_{\emptyset})$, the only $\sim_{\emptyset}$-loops are of size 2 because we have restricted to one-step relations and so have avoided cycles. To ensure that the Herwig extension will still be a tree, we need to appeal to a more general version of Herwig's theorem \cite{Herwig1998ExtendingPI}. Now follow some definitions needed for its formulation.

\begin{definition}  Let $\sigma$ be a relational language. A $\sigma$-structure $L$ is a \textit{link structure} if there are $a_1,...,a_k\in L$ and $R\in\sigma$ such that $|L|=\{a_1,...,a_k\}$ and $L\models Ra_1,..,a_k$ or if $|L|$ is a singleton. Let $\mathcal{L}$ be a set of link structures (all in the language $\sigma$) and $M$ be a $\sigma$-structure. $M$ \textit{is of link type} $\mathcal{L}$ if for every substructure $M'$ of $M$: if $M'$ is a link structure, then there exist $L \in\mathcal{L}$ such that $M'\cong L$ are isomorphic (as $\sigma$-structures). $M$ is called a \textit{packed} structure if for every pair of distinct $a_0,a_1\in M$ there exists a link structure $L$ with $L\subseteq M$ (i.e. $L$ is a substructure\footnote{We say that $L$ is a substructure of $M$ if the inclusion of the domains $i:L\to M$ is an embedding.} of $M$) with $a_1,a_2\in L$. A structure $M$ is called \textit{irreflexive}, if for every $k\in\omega$, every $k$-ary $R$ in $\sigma$ and every $a_1,...,a_k\in M$, if $Ra_1...a_k$ holds, then $a_1,...,a_k$ are pairwise distinct. Let $\mathfrak{F}$ be a set of finite $\sigma$-structures. Let $M$ be a $\sigma$-structure. Finally, we say that $M$ is $\mathfrak{F}$-\textit{free} if there does \textit{not} exist $F\in\mathfrak{F}$ and a homomorphism $\rho:F\to M$.
\end{definition}

\begin{theorem}{\textbf{General Herwig}}\\
Let $\sigma$ be a finite relational language, and $\mathcal{L}$ a set of irreflexive link structures, $\mathfrak{F}$ a set of finite, irreflexive, packed structures, all in the language $\sigma$. Let $C$ a finite $\mathfrak{F}$-free $\sigma$-structure of link type $\mathcal{L}$ with $\{p_1,...,p_k\}$ a (finite) set of partial isomorphisms on $C$. Then there exists a \textit{finite} extension $C^+$ of $C$ of link type $\mathcal{L}$ that is $\mathfrak{F}$-free and satisfies the following conditions:
\begin{itemize}
    \item [(i)] Every $p_i$ extends to a unique automorphism $\widehat{p_i}$ of $C^+$. This yields a subgroup $\langle\widehat{p_1},...,\widehat{p_k}\rangle$ of the automorphism group of $C^+$.
    \item [(ii)] If a tuple $\mathbf{a}$ from $C^+$ is either live or a singleton, then there exists an automorphism $f\in\langle\widehat{p_1},...,\widehat{p_k}\rangle$ such that $f(\mathbf{a})$ lies completely in $C$.
    \item [(iii)] If $\exists f\in\langle\widehat{p_1},...,\widehat{p_k}\rangle$ and $a,b\in C$ such that $f(a)=b$, then either $f=id$ or there is a unique $p\in\langle p_1,...,p_k\rangle$ such that $\widehat{p}=f$ and $p(a)=b$.
\end{itemize}
\end{theorem}

Delete all reflexive arrows for all relations $\sim_X^1$ from $\mathbb{A}_{cut}$ and let $\mathcal{L}:=\{L\subseteq\mathbb{A}_{cut}\;|\;L\;\textrm{a link structure}\}$. Then by definition $\mathbb{A}_{cut}$ is of link type $\mathcal{L}$ and $\mathcal{L}$ is irreflexive. Let $\mathfrak{F}:=\{C_m\;|\;m>2\}$\footnote{Because $\sim_{\emptyset}^1$ is symmetric, we allow for cycles or loops that move to a one-step successor and then immediately back again.}, where $C_m$ is the $m$-sized $\sim_{\emptyset}$-cycle, i.e. where $|C_m|:=\{c_1,...,c_m\}$ and $I(\sim_{\emptyset}^1):=\{(c_i,c_j)\;|\;i\ne j\}$. Clearly, each $C_m$ is finite, packed, and irreflexive. Moreover, we observed that $\mathbb{A}_{cut}$ has no $\sim_{\emptyset}$-loops of size $>2$, hence it is $\mathfrak{F}$-free. Again, for each $h\in A_{cut}$ of $lh(h)=3$ and $last(h)=b\in\mathbb{B}$, there is a \textit{unique} partial isomorphism $p_{h}:h\mapsto h_b$, where $h_{b}:=(b_0,\emptyset,b)$ is of length 2. Set $\{p_{h}\;|\;lh(h)=3\}=\{p_1,...,p_k\}$. \newline\newline
By general Herwig's theorem, there exists a finite $\mathfrak{F}$-free extension $\mathbb{A}_{cut}^+:=(A_{cut}^+,\sim_X^1,D_Xy,P\mathbf{x})$ of link type $\mathcal{L}$ satisfying conditions (i)-(iii) w.r.t. $\{p_1,...,p_k\}$. W.l.o.g., we assume that $\mathbb{A}_{cut}^+$ is a generated submodel, so that $(A_{cut}^+,\sim^1_{\emptyset})$ is connected. Furthermore, by the theorem it is $\mathfrak{F}$-free and hence acyclic up to simply cycles of size $2$. it follows that for any two states $s,t\in A_{cut}^+$ there exists a \textit{unique shortest path} $(s=s_0,...,s_n=t)$ such that $s_0\sim_{\emptyset}^1s_1\;...\;s_{n-1}\sim_{\emptyset}^1s_n$.\footnote{Though the structure is strictly speaking not a tree.} Now we define the new relations as a suitable transitive closure of the one-step relations along these unique shortest paths:
\begin{align*}
    s\sim_X^{tr}t\qquad\textrm{iff}&\qquad\;s_0\sim_X^1....\sim_X^1s_n,\;\textrm{where}\;(s=s_0,...,s_n=t)\;\textrm{is the unique}\;\\
    &\qquad \;\textrm{shortest path between}\;s\;\textrm{and}\;t,\;\textrm{or}\;s=t
\end{align*}
now consider the Herwig extension $\mathbb{A}_{cut}^+:=(A_{cut}^+,\sim_X^{tr}, D_Xy, P\mathbf{x})$ with the new relations $\sim_X^{tr}$ and with the same interpretations for unary predicates of the form $D_Xy(\;)$ or $P\mathbf{x}(\;)$.

\begin{proposition}
$\mathbb{A}_{cut}^+:=(A_{cut}^+,\sim_X^{tr}, D_Xy, P\mathbf{x})$ is a standard relational model. 
\begin{proof}
(1) is an easy consequence of the definition of $\sim_X^{tr}$, together with the fact that the shortest path between $s$ and $u$ is always composed of the transitions from either the shortest path from $s$ to $t$ or the shortest path from $t$ to $u$. By conditions (i) and (ii), we know that for every $s$ there is some automorphism $f\in\langle\widehat{p_1},...,\widehat{p_k}\rangle$ such that $f(s)=h\in A_{cut}$ where $lh(h)\leq 2$. It follows that $tp_{\mathbb{A}_{cut}^+}(s)=\{\phi(v)\;|\;\mathbb{A}_{cut}^+\models\phi(s)\}=tp_{\mathbb{A}_{cut}^+}(h)$. Using the fact that the cut-off model satisfies (2) and an analogue of the restricted truth lemma (lemma 3.1), conditions (2),(3) and (7) follow. \newline\newline
For (4), let $s\sim_X^{tr}t$ and suppose that $P\mathbf{y}(s)$ holds, where $\mathbf{y}=(y_1,...,y_r)$ with $\{y_1,...,y_r\}\subseteq X$ and $(s=s_0,...,s_n=t)$ is the unique shortest path between $s$ and $t$. By induction, we can show that each one-step transition carries over truth of the formula $P\mathbf{y}$ using condition (ii) on live tuples of the form $(s_i,s_{i+1})$. Condition (5) holds because we took a generated submodel, and finally (6) follows from uniqueness of the shortest path.
\end{proof}
\end{proposition}

Our definition of dependence bisimulation also comes completely natural as a notion of bisimulation relation between standard relational models. Taking the exact same relation as before:
\[Z:=\{(s,h)\in A_{cut}^+\times A;|\;\exists f\in\langle\widehat{p_1},...,\widehat{p_k}\rangle\;\textrm{s.t.}\;f(s)=h'\;\textrm{for some}\;h'\in A_{cut}\;\textrm{with}\;last(h)=last(h')\}\]
a similar argument shows that $Z$ is a dependence bisimulation between the standard relational models $\mathbb{A}_{cut}^+$ and $\mathbb{A}$.

\begin{corollary}{\textbf{Modal Finite Model Property}}\\
Every satisfiable $\varphi\in LFD$ has a finite standard relational model.
\end{corollary}

\bibliography{biblio.bib}
\bibliographystyle{eptcs}
\end{document}